\documentclass[twocolumn,amssymb,aps, prd,superscriptaddress,balancelastpage,longbibliography]{revtex4-2}

\newcommand{\env}[1]{\texttt{#1}}
\usepackage{amsfonts}
\usepackage{amsmath}
\usepackage{bm}
\usepackage{graphicx}
\usepackage[breaklinks=true,colorlinks=true,linkcolor=blue,urlcolor=blue,citecolor=blue]{hyperref}
\usepackage{siunitx}
\usepackage{xcolor}
\usepackage{comment}
\usepackage[english]{babel}

\newcommand{\affIQ}{\affiliation{Institut f\"ur Quantenoptik, Leibniz Universit\"at Hannover,\\ Welfengarten 1, D-30167 Hannover, Germany}}

\newcommand{\affSY}{\affiliation{LNE-SYRTE, Observatoire de Paris, Universit\'e PSL, CNRS,\\ Sorbonne Universit\'e 61 avenue de l'Observatoire 75014 Paris}}

\begin{document}

\title[Resolution of the Co-location Problem in Satellite Quantum Tests of the UFF]{Resolution of the Co-Location Problem\\in Satellite Quantum Tests of the Universality of Free Fall}%

\author{Sina Loriani}
\email{loriani@iqo.uni-hannover.de}
\affIQ
\author{Christian Schubert}
\affIQ
\author{Dennis Schlippert}
\affIQ
\author{Wolfgang Ertmer}
\affIQ
\author{\\Franck \surname{Pereira Dos Santos}}
\affSY
\author{Ernst Maria Rasel}
\affIQ
\author{Naceur Gaaloul}
\email{gaaloul@iqo.uni-hannover.de}
\affIQ
\author{Peter Wolf}
\email{peter.wolf@obspm.fr}
\affSY

\begin{abstract}
A major challenge common to all Galilean drop tests of the Universality of Free Fall (UFF) is the required control over the initial kinematics of the two test masses upon release due to coupling to gravity gradients and rotations. 
In this work, we present a two-fold mitigation strategy to significantly alleviate the source preparation requirements in space-borne quantum tests of the UFF, using a compensation mechanism together with signal demodulation. To this end, we propose a scheme to reduce the gravity-gradient-induced uncertainties in an atom-interferometric experiment in a dedicated satellite mission and assess the experimental feasibility. 
We find that with moderate parameters, the requirements on the initial kinematics of the two masses can be relaxed by five orders of magnitude. This does not only imply a significantly reduced mission time but also allows to reduce the differential acceleration uncertainty caused by co-location imperfections below the $10^{-18}$ level.
\end{abstract}

\maketitle
\renewcommand{\vec}[1]{\mathbf{#1}}
\newcommand{\Oo}{\Omega_\text{orbit}}
\newcommand{\Oy}{\Omega_y}
\newcommand{\Om}{\Omega_m}
\newcommand{\keff}{k_\text{eff}}
\newcommand{\Tzz}{T_{zz}^0}
\newcommand{\dt}{~ \text{d}t}
\newcommand{\Dijk}{\Delta_k^{i,j}}
\newcommand{\dO}{{\delta\Omega}}
\newcommand{\vkeff}{\vec{k_\text{eff}}}

\newcommand{\Eotvos}{E\"otv\"os}

%==============================================================
\section{Introduction}\label{sec:intro}
The Equivalence Principle is a remarkable concept of physics as it threads its way through scientific history, facilitating our understanding of gravity since the times of Galileo and Newton. 
Postulating the equivalence of inertial and gravitational mass implies the same free fall acceleration of objects of different composition, which has been labeled as the Weak Equivalence Principle or Universality of Free Fall (UFF).
This notion, together with the principle of Relativity, today lays the foundation for General Relativity (GR), which constitutes the present perception of the macroscopic world. 
Even more, in its modern formulation comprising the UFF, Local Lorentz Invariance and Local Position Invariance, the Einstein Equivalence Principle (EEP) consolidates the assumptions required to comprehend gravity as a purely geometrical phenomenon and therefore serves as classification for gravitational theories \cite{Will14}. Gravity is however the only fundamental force of nature that could not yet be integrated into the Standard Model, which explains particle phenomena on the microscopic scale with outstanding success, ranging from high-energy physics as observed in particle colliders to the ultra-cold realm of atom optics. 
Moreover, the significance of Dark Energy and Dark Matter for cosmological considerations supports the strive to unveil a more fundamental, general theory that yields General Relativity and the Standard Model as low energy limits.
Attempts to find such a theory predict a violation of the EEP by introducing additional forces or fields that break the universal coupling of gravity to matter \cite{Kostelecky11,Damour12}.
As a consequence, despite its elegant simplicity and its hitherto unchallenged success, the Equivalence Principle is subject to a large variation of validation tests including, for example, tests of the gravitational redshift of clocks, or of local Lorentz Invariance.  
Among those experiments, special attention is paid to the UFF, as Schiff's conjecture \cite{Schiff60} and arguments based on energy conservation \cite{Nordtvedt75} indicate that violation of one of the constituents of EEP implies a violation of the others, and UFF tests are likely to be the most promising route to detect such a violation.

In experiments searching for a UFF violation, the figure of merit is given by the \Eotvos\ parameter $\eta = \Delta a/g$ which quantifies the differential acceleration $\Delta a = \vec n \cdot (\vec a_A- \vec a_B)$ of two test masses $A$ and $B$. The sensitive axis $\vec n$ denotes the direction along which the local gradient $g = \vec n \cdot \vec g$ of the  gravitational field is measured. 
To date, all experiments have confirmed the UFF, corresponding to $\eta = 0$, with ever-increasing accuracies, which lie at $\delta \eta \sim 10^{-13}$ - $10^{-14}$ \cite{Wagner12,Hofmann18,Touboul17}.
As a rather recent development, inertial-sensitive matter wave interferometry opened up a new pathway in testing the UFF by comparing the gravitation-induced phase shift for two different, freely falling matter waves.
As such, they belong to the class of Galilean drop tests, as opposed to force balance experiments, and significantly extend the set of test-mass pairs to a wide range of atomic species. 
This is of great importance in constraining various composition-dependent violation scenarios such as dilaton models \cite{Damour12} motivated by String theory and  parametrized frameworks such as the Standard Model Extension \cite{Kostelecky11,Hohensee13}. 
Moreover, the coupling of gravity to matter can be investigated on a quantum-mechanical level by introducing spin degrees of freedom \cite{Laemmerzahl06,Tarallo14}, superposition of electronic states \cite{Rosi17} and by studying the effect of gravity onto the internal dynamics \cite{Pikovski15,Roura20,Ufrecht20}.
So far, the UFF has been tested in the $10^{-7}$-$10^{-12}$-range \cite{Fray04,Bonnin13,Schlippert14,Tarallo14,Rosi17,Zhou19arxiv,Albers20,Asenbaum20arxiv} in different atom interferometry setups with various isotopes and elements.
Since the sensitivity scales with the free fall time of the atoms, large atomic fountain experiments and space-borne missions predict accuracies in the $10^{-15}$ regime \cite{Aguilera14,Hartwig15,Williams16,Overstreet18} and beyond \cite{Berge19}, competing with the best classical tests \cite{Wagner12,Hofmann18,Touboul17}.

It is, however, well known that the accuracy of drop tests is limited by the preparation of the two sources \cite{Blaser01}.
Indeed, any deviation from a uniform gravitational field leads to an acceleration that depends on the initial coordinates of a test mass, that is its initial position $\vec r_0$ and velocity $\vec v_0$, irrespective of whether that test mass is macroscopic or a matter wave.
In particular, gravity gradients $\Gamma$, the second order derivative of the local gravitational field, give rise to a spurious (time-dependent) differential acceleration
\begin{equation}\label{eq:a_GG}
    \Delta \vec a_{GG} = \Gamma \left(\Delta \vec r_0 + \Delta \vec v_0 t \right)~,
\end{equation}
which, a priori, can not be distinguished from the linear acceleration that is to be measured. 
Consequently, in an experiment searching for minuscule violations of the UFF, the initial co-location of the two test masses in position $\Delta \vec r_0=\vec r_{0,A}-\vec r_{0,B}$ and velocity $\Delta \vec v_0=\vec v_{0,A}-\vec v_{0,B}$ has to be accurately determined, since uncertainties in the initial kinematics directly translate into a systematic uncertainty $\delta \Delta \vec a_{GG} = \Gamma \left(\delta \Delta \vec r_0+\delta \Delta \vec v_0 t\right)$ in the measurement of the differential acceleration $\Delta a$~\footnote{
    Throughout this paper, we will use $\Delta$ and $\delta$ to denote a difference and an uncertainty, respectively. For example, $\Delta \vec r_0 = \vec r_{0,A}-\vec r_{0,B}$ is the displacement of species $A$ and $B$, and $\delta \Delta \vec r_0$ denotes the uncertainty in that quantity.
}.

In quantum tests of the UFF, the test masses are two carefully prepared wave packets. 
In phase space, these quantum states follow statistical distributions around experimentally realized means.
Due to their statistical nature, a certain number $\nu$ of realizations is required in order to determine the mean differential position and velocity within desired uncertainty $\delta \Delta \vec r_0$ and $\delta\Delta \vec v_0$, given by
\begin{equation}\label{eq:ver}
\delta \Delta  r_{0,i} = \frac{\sigma_{r,0,i}}{\sqrt{\nu N/2}} 
\hspace{0.5cm} \text{and} \hspace{0.5cm} 
\delta \Delta v_{0,i} = \frac{\sigma_{v,0,i}}{\sqrt{\nu N/2}},
\end{equation}
where $\sigma_{r,0,i}$ and $\sigma_{v,0,i}$ denote the spatial extent and velocity width of one atomic ensemble, respectively~\footnote{
    $\delta\Delta \vec r_{0,i} = \left(\sigma_{r,0,i,A}^2/N_A+\sigma_{r,0,i,B}^2/N_B\right)^{1/2}/\sqrt{\nu}$. For simplicity, we assume similar values for both clouds, $\sigma_{r,0,i,A}=\sigma_{r,0,i,B}=\sigma_{r,0,i}$ and $N_A=N_B=N$.
}. 
$N$ is the number of atoms in the atomic sample and $i=x,y,z$ denotes the spatial coordinate. 
One realization corresponds to imaging the atomic cloud in situ or after time-of-flight to infer spatial or velocity-related properties, respectively.
Given that the number $N$ of atoms per shot is limited and that the product of the sizes $\sigma_{r,0,i}$ and $\sigma_{v,0,i}$ is fundamentally constrained by Heisenberg's principle, the number $\nu$ of required verification shots can be fairly high and make up a large part of a measurement campaign.
As an example, the uncertainty in the differential mean position $\delta \Delta \vec r_0$ of the two test masses has to be determined to the nm level to keep the effect of \eqref{eq:a_GG} below $\delta \eta = 10^{-15}$ in a space-borne UFF test \cite{Aguilera14}. For an atom interferometer with typical experimental parameters, this requires $\nu\sim 10^5$ shots with $N=10^6$ atoms. In view of this unfavorable scaling, considering even more ambitious scenarios targeting $\delta \eta = 10^{-17}$ is futile, as the displacement would need to be controlled at the \SI{10}{pm} level. 

However, these long integration times can be avoided by artificially introducing accelerations that compensate the gravity gradient induced acceleration \eqref{eq:a_GG} and hence alleviate the dependency on the initial preparation, as proposed in \cite{Roura17} and already implemented in ground-based experiments \cite{Overstreet18,Damico17,Asenbaum20arxiv}.
In this work, we generalize this compensation technique to space-borne missions with time-dependent gravity gradients, and study its feasibility in combination with signal demodulation, in which one takes advantage of the spectral separation between the target signal and the gravity-gradient-induced perturbation \cite{Touboul17}.
With this two-fold strategy, the determination of the initial position (velocity) to the $\mu$m ($\mu$m/s) level is sufficient, compatible with state-of-the-art laboratory capabilities, such that only a few verification shots $\nu$ are required. Even more, this allows to integrate gravity gradient induced acceleration uncertainties below the $10^{-18}$ level in atom-interferometric tests of the UFF within favorable experimental parameter scales.

\section{Gravity gradient compensation}\label{sec:ggc}
\subsection{Model}\label{ssec:ggc_model}
The Mach-Zehnder configuration \cite{Kasevich91} is the most common atom interferometer geometry for inertial applications.
A beam-splitter ($\pi/2$) light grating creates a coherent superposition of momentum states, which propagate freely for a duration $T$ before being redirected by a mirror ($\pi$) pulse such that after an equal propagation time $T$, a final $\pi/2$ beam-splitter recombines the two wave packets.
The two output ports of the interferometer differ in momentum, and their relative population is a function of the accumulated differential phase $\phi$ between the two interferometer branches.
In our analysis, we follow a semi-classical description, in which the phase shift is evaluated by inserting the classical trajectories into the phase expression \cite{Storey94,Antoine03,Hogan08}
\begin{equation}\label{eq:phi}
\begin{aligned}
    \phi =\ & \vec r_0\cdot\vec{\keff}^{(1)}\\
    & -2\frac{\vec r_u(T)+\vec r_l(T)}{2}\cdot\vec{\keff}^{(2)}\\
    & +\frac{\vec r_u(2T)+\vec r_l(2T)}{2}\cdot\vec{\keff}^{(3)}
\end{aligned}
\end{equation}
for a Mach-Zehnder configuration. Here, $\vkeff^{(j)}$ is the wave vector of the $j^\text{th}$ light pulse ($j=1,2,3$), and $\vec r_{u}$ ($\vec r_{l}$) the classical position of the wave packet on the upper (lower) branch of the interferometer upon interaction with the light in a coordinate system tied to the satellite frame. 
Typically, the three pulses $\vkeff^{(j)} = \vkeff = \keff\,\vec n$ are identical, where $\vec n$ indicates the sensitive axis of the interferometer.
The projection of the atoms' free fall acceleration $\vec a$ on this axis gives rise to the leading order phase shift, $\phi_a = \keff\,\vec{n}\cdot\vec a\,T^2$, which allows to directly assess the \Eotvos\ parameter $\eta$ in a differential measurement.
\begin{figure*}
    \centering
    \includegraphics[width=\textwidth]{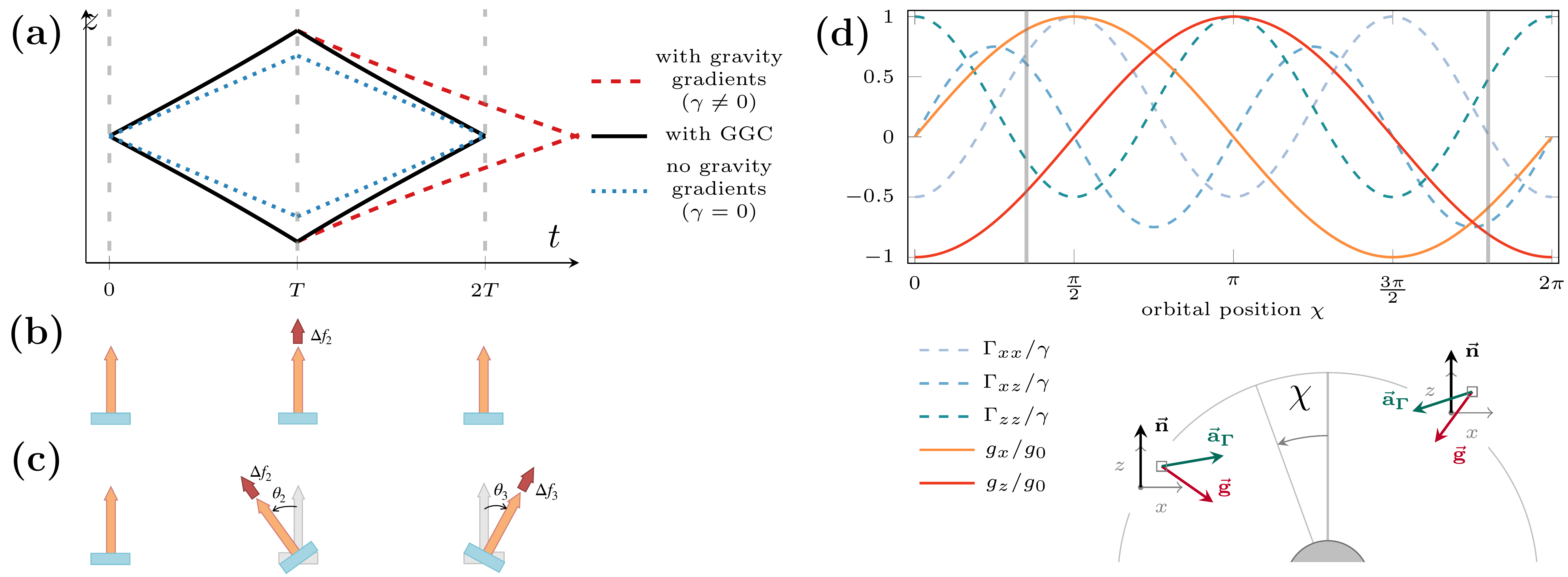}
    \caption{\textbf{Gravity gradient compensation in a space mission.} (\textbf{a}) In the presence of gravity gradients, the straight trajectories (blue, dashed) of the atoms in a freely falling frame get deformed, leading to an open (red, long dashes) interferometer. It is closed (solid, black) through (\textbf{b}) application of an appropriate frequency shift \cite{Roura17} at the second pulse. (\textbf{c}) In this work, the gradient compensation technique is  extended to two dimensions by tilting the laser and changing it in frequency at the second as well as at the third light pulse. (\textbf{d}) This is required to mitigate the varying local values of the gravity gradient tensor components in the satellite frame in an inertial space mission. The effective acceleration $\vec a_\Gamma$ due to gradients and due to linear gravitational acceleration $\vec g$ shall only depict the changes in direction over an orbit and are not to scale. The sensitive axis and the satellite position on the orbit are labelled by $\vec n$ and $\chi$ respectively. \label{fig:ggc_scheme}}
\end{figure*}
This treatment is exact for Lagrangians up to quadratic order in position and velocity, hence serving the purpose to study the effects related to gravity gradients (see Appendix~\ref{app:model} for details).
The duration of atom-light-interaction $\tau$ is assumed to be small compared to the pulse separation time $T$, which is the case for space-borne experiments with long drift times on the order of seconds. However, the treatment can be extended to account for pulses of finite duration leading to corrections in the order of $\tau/T$ \cite{Bertoldi19,Wang18}.

The gravity-gradient compensation (GCC) technique proposed in \cite{Roura17} exploits that the gravity gradients introduce phase shifts  $\phi_{GG} = \vkeff\cdot\vec a_{GG} T^2$ (see Eq.~\eqref{eq:a_GG}), which linearly depend on the initial position and velocity of an atom and may be compensated by introducing a controllable shift with similar dependency. 
Indeed, the phase expression \eqref{eq:phi} features a linear dependency on the atom's position $\vec r$, such that an additional shift at the mirror pulse, $\vkeff^{(2)}=\vkeff+\delta\vkeff$ gives rise to terms proportional to $\delta \vkeff$ and the initial coordinates of the atom. 
In another picture, this corresponds to closing the interferometer deformed by gravity gradients as depicted in Fig.~\ref{fig:ggc_scheme}a. 
It is interesting to note that also higher orders of the gravitational potential (cubic and higher) can be compensated in a similar fashion, as can be shown in a perturbative treatment \cite{Bertoldi19, Ufrecht20b}.

\begin{comment}
\begin{figure*}
    \centering
    \includegraphics[width=\textwidth]{figs_and_tabs/ggc.pdf}
    \caption{\textbf{Gravity gradient compensation in a space mission.} (\textbf{a}) In the presence of gravity gradients, the straight trajectories (green, dashed) of the atoms in a freely falling frame get deformed, leading to an open (red, long dashes) interferometer. (\textbf{c}) It is closed (solid, black) through application of an appropriate frequency shift \cite{Roura17} at the second pulse. (\textbf{c}) In general, the local gravity gradients are varying in a space mission, denoted by the changing amplitude and sign of the gravity gradient tensor components. (\textbf{d}) The gradient compensation technique is hence extended to two dimensions by tilting the laser and changing it in frequency at the second as well as at the third light pulse. \label{fig:ggc_scheme}}
\end{figure*}
\end{comment}
Anticipating the application to satellite missions, in which the gravity gradients are temporally varying and couple to rotations of the apparatus, we generalize this idea to the wave vectors
\begin{equation}\label{eq:keff}
\begin{aligned}
\vec \keff^{(j)}=
        \begin{pmatrix} k^{(j)}_{\text{eff},x} \\ 
            k^{(j)}_{\text{eff},y} 
            \\ k^{(j)}_{\text{eff},z}
        \end{pmatrix} 
      = \begin{pmatrix} \keff \Delta_{x,j} \\
           0
            \\ \keff(1+\Delta_{z,j})
        \end{pmatrix} \\
\end{aligned}
\end{equation}
for each pulse by introducing controllable shifts $\Delta_{x,j}$ and $\Delta_{z,j}$ for $j=2,3$, with $\vkeff^{(1)} = (0,0,\keff) =: \vkeff$. 
In an experiment, realizing these wave vectors corresponds to shifting the laser in frequency and tilting it relative to the first pulse, as detailed in Appendix~\ref{app:realization} and illustrated in Fig.~\ref{fig:ggc_scheme}b and c.
For more general applications, one might introduce additional shifts in the $y$-direction.
However, we will focus on satellites that spin in the orbital plane which we set to coincide with the $x$-$z$-plane.

In the satellite frame, the Lagrangian describing the free fall of an atom may be written as
\begin{equation}\label{eq:L}
L = \frac 1 2 m\left( \dot{\vec{r}} + \vec\Omega_s\times\vec r \right)^2 + m \vec a(t)\vec r + \frac 1 2 m \vec r\Gamma(t)\vec r,
\end{equation}
where $\vec \Omega_s$ accounts for the spinning of the satellite, $m$ is the atomic mass and $\Gamma(t)$ denotes the local gravity gradient tensor.
Note that under the assumption of the UFF, the Lagrangian is independent of the linear gravitational acceleration in an inertial reference frame.
The acceleration term $\vec a(t)$ needs however to be included in the treatment since it comprises the sensitivity to a possible UFF violation $\eta \vec g(t) = \vec a_A(t)-\vec a_B(t)$ in a differential measurement of two species $A$ and $B$.

The interferometer phase is obtained by solving the (classical) equations of motion for segment-wise freely falling atoms, with boundary conditions defined by the wave vectors \eqref{eq:keff}.
The solution is obtained by virtue of a power-series ansatz \cite{Hogan08} for the trajectories. Then, using the Lagrangian \eqref{eq:L} the phase can be written as
\begin{equation}\label{eq:phi_dep}
        \phi =  \phi_\text{indep.} +\sum^3_{i=1}\alpha_i r_{0,i} + \sum^3_{i=1}\beta_i v_{0,i}
\end{equation}
by collecting the dependencies on the initial position $r_{0,i}$ and velocity $v_{0,i}$ in the coefficients $\alpha_i$ and $\beta_i$, respectively, with $i=x,y,z$.
$\phi_\text{indep}$ comprises all contributions that are independent of the initial conditions.
The coefficients $\alpha_i$ and $\beta_i$ are, among other experimental parameters, functions of the wave vector shifts $\Delta_{i,j}$ introduced in Eq.~\eqref{eq:keff}.
Therefore, the unwanted phase dependencies on the initial kinematics are compensated by requiring $\alpha_i = \beta_i = 0$, which yields explicit expressions for $\Delta_{i,j}$.

%==============================================================
\subsection{Results}\label{ssec:ggc_results}
In the case of a stationary ground experiment, in which, to leading order, the gradient tensor is given by $\Gamma=\text{diag}(-\gamma/2,\gamma/2,\gamma)$ with $\gamma=2GM_E/R_E^2$ (with $M_E$, $R_E$ being Earth's mass and radius, respectively, and $G$ the gravitational constant), we indeed recover the result $\Delta_{z,2}=\gamma T^2/2$ (the other shifts being zero) of reference \cite{Roura17} when neglecting rotations, $\vec \Omega_s = 0$.
Similarly, for $\Gamma= 0$ and $\vec \Omega_s = (0,\Omega_y,0)$, we find $\Delta_{x,2} = - \sin(\Omega_y T)$, $\Delta_{z,2} = -1 +\cos(\Omega_y T)$, $\Delta_{x,2} = - \sin(2\Omega_y T)$ and $\Delta_{z,2} = -1 +\cos(2 \Omega_y T)$. This corresponds to counter-rotating the laser (mirror) between two pulses by the angle $\Omega_y T$ to compensate for rotations, a well known result used in ground-based experiments to account for Coriolis forces introduced by the rotation of the Earth \cite{Lan12}.

\begin{figure}
    \centering
    \includegraphics[width=\columnwidth]{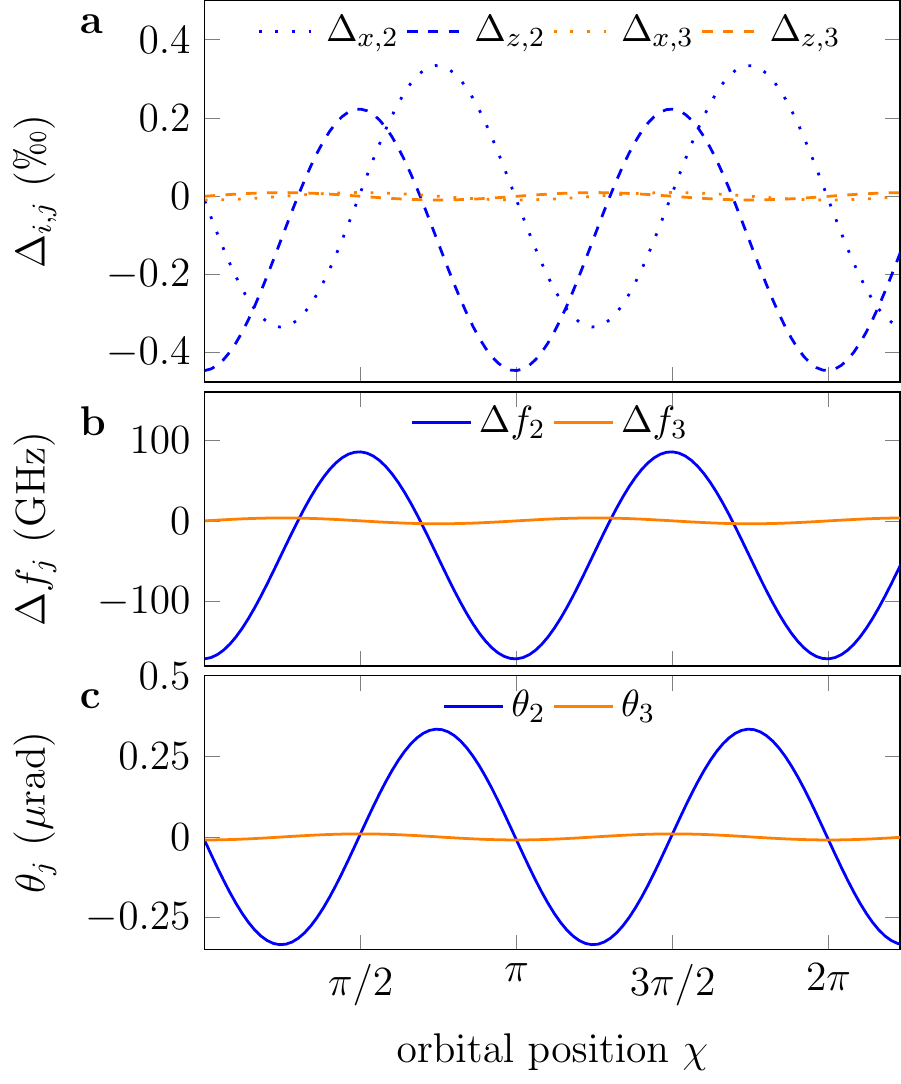}
    \caption{\textbf{Parameters for gravity gradient compensation in an inertial satellite mission. a)} Fractional momentum vector shifts in $x$ (dotted) and $z$ (dashed) direction. These shifts are realized by periodically \textbf{b)} shifting the laser in frequency and \textbf{c)} tilting the setup with respect to the first pulse. In all plots, blue (orange) corresponds to the value at the second (third) pulse, and $\chi$ denotes the orbital position when the first pulse is applied. The assumed parameters target $\delta \eta \leq 10^{-17}$ and are stated in Tab.~\ref{tab:params}. \label{fig:shifts}
    }
\end{figure}

In this study, we focus on the case of a satellite in inertial configuration (i.e. it keeps its orientation with respect to a celestial reference system, $\vec \Omega_s = 0$) on a circular orbit. The effect of residual rotations $\delta \vec \Omega\neq 0$, however, is taken into account in the error assessment in Sec.~\ref{sec:uff}. 
The assumed spherically symmetric gravitational potential of the Earth allows for an analytical calculation. The concepts of this paper, however, can be extended to arbitrary orbits and more sophisticated gravitational potential models in a numerical treatment. 
An important feature in the system under consideration is the modulation of the gravitational field components in the local frame of the satellite as illustrated in Fig.~\ref{fig:ggc_scheme}d and detailed in Appendix~\ref{app:model}. 
In particular, the values of the gravity gradient tensor are modulated at twice the orbital frequency, which is $\Oo = \sqrt{GM_E/(R_E+h_\text{sat})^3}$  for a circular orbit at altitude $h_\text{sat}$. 
As a consequence, the required compensation shifts $\Delta_{i,j}$ have to be modulated in a similar fashion, as displayed in Fig.~\ref{fig:shifts}. Their magnitude is mainly determined by the scale factor $\keff T^2$ of the interferometer and the value of the local gravity gradients, and they are given by
\begin{equation}\label{eq:sol}
    \begin{aligned}
        \Delta_{x,2}&= \frac 3 8 \gamma T^2\sin(2\chi)+ \frac 5 8 \gamma T^3\Oo\cos(2\chi)+...\\
        \Delta_{z,2}&= \frac 1 8 \gamma T^2 \left(1+3\cos(2\chi)-5T\Oo \sin(2\chi)\right)+... \\
        \Delta_{x,3}&= \frac 1 2 \gamma T^3 \Oo \cos(2\chi)+... \\ 
        \Delta_{z,3}&= -\frac 1 2 \gamma T^3 \Oo \sin(2\chi)+... 
    \end{aligned}
\end{equation}
to first order in $\gamma T^2$, where $\chi$ is the angle characterizing the orbital position (see Fig.~\ref{fig:ggc_scheme}b).

\section{Signal demodulation}\label{sec:demodulation}
A decisive advantage of space tests of the UFF is the inherent modulation of the signal. As alluded to in the previous section, the different components of the gravitational field are modulated at different frequencies. 
The measured differential acceleration signal can hence be decomposed into its frequency components,
\begin{equation}\label{eq:acc_freq}
     \Delta a = \eta g_0 \cos(\Om t)+\Delta a_\text{const}+\sum_{k=1} \Delta a_\text{sys}^k \cos(k\Om t).
\end{equation}
Here, $\eta g_0$ is the differential acceleration introduced by a possible violation, modulated at a certain frequency $\Omega_m$. 
This frequency corresponds to the orbital frequency $\Oo$ for inertial configurations, or to $\Om = \Oo+\Omega_s$ for a satellite spinning in the orbital plane by $\Omega_s$. All non-varying contributions or very slow drifts (i.e. varying on time scales $\ll 2\pi/\Omega_m$) are comprised in $\Delta a_\text{const}$, for example a differential acceleration caused by a constant magnetic field bias. Finally, we consider systematic contributions $\Delta a_\text{sys}^j$ at higher harmonics $j\Om$ of the modulation, with gravity gradients varying at $2\Om$. As in the previous section, we suppose a simplified scenario with a circular orbit and a spherical gravitational potential for clarity. The following considerations can, however, be extended to continuous frequency spectra \cite{Touboul17}.  

Demodulation of the differential acceleration signal at the target frequency $\Om$, at which the violation is expected, for a duration $\tau$ is given by
\begin{widetext}
\begin{equation}\label{eq:avg}
\begin{aligned}
    & \frac{2}{\tau} \int_0^\tau \Delta a \cos(\Om t)\dt = (\eta g_0+\Delta a_{sys}^1) \\
    &+\frac {2}{\tau \Omega_m} \left[\frac {\eta g_0+\Delta  a_{sys}^1}{2}\sin (2\Omega_m \tau)+\Delta a_{const}\sin(\Omega_m \tau)+ 2\sum_{k=2}\Delta a_{sys}^k\left[\frac{\sin([k\Omega_m-\Omega_m]\tau)}{k\Omega_m-\Omega_m}+\frac{\sin([k\Omega_m+\Omega_m]\tau)}{k\Omega_m+\Omega_m}\right] \right]\\
    &\leq \left(\eta g_0+\Delta a_\text{sys}^1 \right) + \frac{2}{\tau \Om}\left(\frac{\eta g_0}{2}+\left|\Delta a_\text{const}\right|+\frac 4 3\sum_{k=2}\left|\Delta a_\text{sys}^{(k)}\right|\right),
\end{aligned}
\end{equation}
\end{widetext}
where $\Delta a_\text{sys}^1$ displays any components of the systematics (only co-location related effects in the scope of this paper), which are modulated at the same frequency as a possible violation signal. In the scenario under consideration, for an inertial mission on a perfectly circular orbit, this contribution is zero. However, any finite ellipticity introduces such a frequency component, as is shown in Appendix~\ref{app:model}.

The final expression shows that the potential violation signal is demodulated to DC, while the contributions at other frequencies and constant terms are integrated down. Here, the modulation frequency $\Om$ determines the rate of integration. With respect to the integration behaviour it may hence be beneficial to spin the satellite in the orbital plane, as for example employed in \cite{Touboul17} and along the lines of \cite{Williams16}. However, spinning the satellite introduces fictitious forces which couple to the initial conditions, too. It is possible to compensate them by counter-rotating the mirror \cite{Lan12} by the angle $T\Omega_s$ between two subsequent pulses. This rotation is additionally modulated with the periodic tilt determined in the previous section for gravity gradient compensation.
Note that the authors of \cite{Chiow17}, too, exploit the fact that the gravity gradients are modulated at a different rate than the gravitational acceleration by introducing an artificial modulation by rotating the experimental setup on a gimbal mount. We find, however, that an additional spinning is not required, even for the ambitious scenario under consideration, as will be demonstrated in the next section.
Finally, the described integration behaviour displays the worst case scenario, as the final expression \eqref{eq:avg} is obtained by taking the upper bound of the trigonometric functions in the intermediate step. In fact, the choice of an adequate integration time $\tau$ allows to evaluate the signal more efficiently by matching the minima of the expression (c.f. minima in Fig.~\ref{fig:integration_curve}).

\section{UFF test scenario}\label{sec:uff}
\subsection{Sensitivity to UFF violations}
The concurrent operation of two matter-wave interferometers employing different atomic species $A$ and $B$ allows to infer the differential acceleration by simultaneous, individual phase measurements $\phi_\alpha = \vec k_{\text{eff},\alpha} a_\alpha T_\alpha^2 + \phi_{\text{sys},\alpha}$ with $\alpha=A/B$. The single-shot quantum projection noise (atomic shot noise)
\begin{equation}
        \sigma_{\Delta a}^{(1)} = %\left[
        \sqrt{\left(\frac{1}{C_A k_{\text{eff},A} T_A^2 \sqrt{N_A}}\right)^2 +\left(\frac{1}{C_B k_{\text{eff},B} T_B^2 \sqrt{N_B}}\right)^2}%\right]^{1/2}
\end{equation}
given by the number $N_\alpha$ of atoms contributing to the signal, is the intrinsic differential acceleration uncertainty per experimental cycle. The contrast $C_\alpha$ accounts for the visibility of the interference fringes. Such a setup is sensitive to violations of the UFF, with the fundamental statistical uncertainty of the \Eotvos-parameter
\begin{equation}\label{eq:sigeta}
    \sigma_\eta = \frac{\sigma_{\Delta a}^{(1)}\sqrt 2}{g_0 \sqrt n}.
\end{equation}
after $n\gg1$ measurements. As explained in Appendix~\ref{app:realization}, the factor $\sqrt 2$ accounts for the sinusoidally varying local value of the gravitational acceleration within a measurement campaign \cite{Schubert13,Berge19}.
In the following discussion, we assume the parameters stated in Tab.~\ref{tab:params} for an exemplary UFF test scenario as presented in \cite{Berge19}, targeting an accuracy of $\delta \eta \leq 10^{-17}$ involving isotopes of rubidium (Rb) and potassium (K). This analysis is not covering all aspects of a mission proposal but rather demonstrates the mitigation of co-location-related systematics for scenarios far beyond the state-of-the-art \cite{Touboul17}. 
Since the free-fall time $T$ in space-borne atom interferometers is not subject to the same limitations as on ground, it can be assumed to be much larger than in table-top experiments or fountains.
Indeed, the coherence and low expansion rate of ultra-cold atomic sources allows to operate on time scales in the order of several seconds \cite{Muentinga13,Aguilera14}.
Due to the geometrical constraints in a satellite mission, the magnitude of momentum transfer $\keff$ is, however, limited, and we choose a second-order double diffraction scheme \cite{Ahlers16} ($\keff = 4 k_L$ with $k_L$ being the laser wave number) in the following, such that the spatial extent of the interferometer is less than $\SI{1}{m}$.  
Moreover, we suppose typical atomic numbers and cycle times for the generation of sufficiently well-engineered quantum sources of Bose-Einstein condensates \cite{Aguilera14,Loriani19}. 
Assuming that 10\,s are required for the atomic source preparation followed by $2T=\SI{40}{s}$ of interferometry, a cycle time of \SI{10}{s} can be achieved supposing an interleaved operation of 5 concurrent interferometers \cite{Savoie18}. Thanks to the choice of modest momentum transfer and the mitigation of major sources of contrast loss, such as gravity gradients, the contrast can be assumed to be near unity. 
With these parameters, the shot-noise limited \Eotvos\ parameter is integrated down to $8\times 10^{-16}$ after one orbit, such that $\sigma_{\eta} \leq 10^{-17}$ can be reached within a total of $\tau = \SI{15}{months}$ of integration, corresponding to $n=4\times10^6$ interferometric measurements.

\begin{table}[t]
\begin{ruledtabular}
    \centering
    \begin{tabular}{cc|c}
         quantity & value & definition \\
         \hline
            $T$                     & \SI{20}{s}     & pulse separation time \\
            $k_\text{eff,Rb}$ & 8$\pi$/(\SI{780}{nm}) & effective wave number (Rb)\\
            $k_\text{eff,K}$ & 8$\pi$/(\SI{767}{nm}) & effective wave number (K)\\
            %$k_{\text{eff,Rb}}$ & \SI{32.2}{$\times 10^6$ m$^{-1}$}  &  effective momentum transfer (rubidium)\\
            %$k_{\text{eff,K}}$ & \SI{32.8}{$\times 10^6$ m$^{-1}$}  &  effective momentum transfer (potassium)\\
            $N$ & $10^6$ & number of atoms per shot\\
            $T_c$ & \SI{10}{s}& cycle time\\
            $\bm{\delta r_{j,0}}$ & \textbf{1~$\bm{\mu}$m}   & \textbf{differential initial position} \\
            $\bm{\delta v_{j,0}}$ & \textbf{1~$\bm{\mu}$m/s}   & \textbf{differential initial velocity}\\
            $h_\text{sat}$ & \SI{700}{km} & orbit height\\
            $\delta \Omega$ & 0.1~$\mu$rad/s & residual satellite rotations \\
            $\delta \gamma$ & $10^{-10}$ s$^{-2}$ & gravity gradient uncertainty\\
            $e$ & $10^{-3}$ & orbit ellipticity\\
            $\delta \theta$& 1~$\mu$rad   & laser tilt angle uncertainty\\
            $\delta f$ & \SI{400}{kHz}     & laser frequency shift uncertainty
    \end{tabular}
    \caption{
   \textbf{Assumed parameters for a UFF test mission on an inertial satellite featuring gravity gradient cancellation and signal demodulation.}
    For the assumed orbit, the maximal value of the gravitational acceleration and gravity gradient tensor are $g_0 = \SI{7.9}{m/s^2}$ and $\gamma= \SI{-2}{\times10^{-6}~s^{-2}}$, respectively, and the orbital frequency is $\Oo=\SI{0.17}{\times 2\pi~\text{mHz}}$. The cycle time $T_c=\SI{10}{s}$ can be realized by the concurrent operation of 5 interferometers, assuming \SI{10}{s} for the preparation of the source. In combination with signal demodulation, the compensation technique allows to reduce the systematic uncertainties linked to gravity gradients by five orders of magnitude for these parameters, which relaxes the requirements on the initial co-location of the species by the same amount. \label{tab:params}}
\end{ruledtabular}
\end{table}

\begin{figure*}
    \centering
    \includegraphics[width=\textwidth]{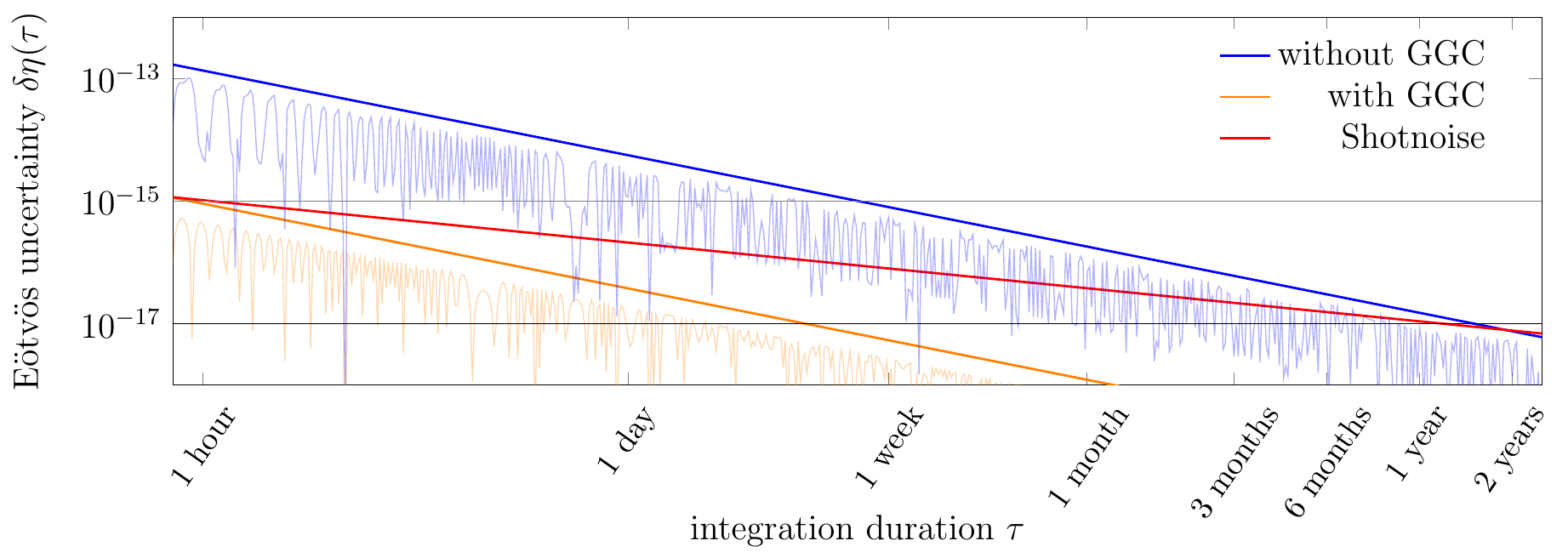}
    \caption{\textbf{Integration of systematic uncertainties due to gravity gradients in a UFF test with Rb and K.} GGC significantly reduces the systematic contributions, such that the residual differential acceleration may be attenuated to unprecedented degree through signal demodulation (orange curve). This does not only allow for largely reduced requirements on the source preparation for mission proposals as STE-QUEST \cite{Aguilera14} but also paves the way for more ambitious mission scenarios \cite{Berge19} targeting $\delta \eta \leq 10^{-17}$ in shot-noise limited operation (red curve). In comparison, although the systematics are integrated down thanks to demodulation, the measurement would be limited by systematics without GGC  (blue curve). \label{fig:integration_curve}}
\end{figure*}

\subsection{Initial kinematics dependence}
As indicated in the introduction, any spurious differential acceleration between the two species can, a priori, not be distinguished from a potential UFF violation signal. Consequently, all systematic error sources have to be controlled at a level better than the target inaccuracy of $\delta \eta = 10^{-17}$, or be modulated at other frequencies than the local projection of $\vec g$. Eq.~\eqref{eq:a_GG} describes how the acceleration of each species is linked to its initial mean position and velocity and constrains the interspecies displacement uncertainty to $\delta \vec r_0 \sim 10$\,pm and $\delta \vec v_0 \sim 1$\,pm/s in position and velocity, respectively. The number of verification measurements $\nu \sim 10^8$ (see Eq.~\eqref{eq:ver}) required to ensure the source preparation at this level would exceed the number of realizations of the actual interferometric experiment by far. Even the less ambitious goal of $\delta \eta = 2\times 10^{-15}$ as in \cite{Aguilera14} would necessitate to allocate a significant part of the mission duration to the analysis of this systematic effect.

However, by employing the recipe outlined in Sec.~\ref{sec:ggc}, we find that for the assumed mission parameters, the gravity gradient induced uncertainties can be compensated by applying the time-dependent momentum vector shifts \eqref{eq:sol} which corresponds to periodically tilting the laser up to $300\,\mu$rad and shifting it in frequency in the order of 150\,GHz as displayed in Fig.~\ref{fig:shifts}. Indeed, the dependencies on the initial kinematics are largely compensated, such that the major residual contributions to the differential acceleration uncertainty stem from:
\begin{itemize}
    \item Imperfections in the experimental realization, mainly given by the tilt error $\delta \theta$:  $\delta\theta \delta x_0/T^2$ and $\delta\theta \delta v_{x,0}/T$  in the order of $\sim  5\times 10^{-14}$~m s$^{-2}$.
    \item Residual satellite rotations $\delta \Omega$: $\delta \Omega \delta v_{x,0}$ and $\delta \Omega \delta v_{y,0}$ in the order of $\sim  10^{-13}$~m s$^{-2}$.
    \item Uncertainties in the knowledge of the local gravity gradient $\delta \gamma$: $\delta \gamma \delta z_{0} \cos(2\chi)$ and $\delta \gamma \delta v_{z,0} T \cos(2\chi)$  in the order of $10^{-16}$~m s$^{-2}$.
\end{itemize}
More details are found in Appendix~\ref{app:realization}. These relations allow for a trade-off between required control of the experimental background ($\delta \Omega, \delta \theta, \delta \gamma$) and characterization of the source preparation ($\delta{\vec r_0}, \delta \vec{v_0}$), leading to the numbers in Tab.~\ref{tab:params}. Note that these numbers are conservative as they stem from a linear (rather than quadratic) sum of uncertainties, although many of those uncertainties are expected to be uncorrelated (c.f. Appendix~\ref{app:realization}).

Most importantly, these contributions are either constant or modulated at twice the orbital frequency, which is the modulation frequency of a potential UFF violation in the given setup. As outlined in Sec.~\ref{sec:demodulation}, this allows to distinguish these accelerations by demodulating the signal, which is of great significance as illustrated in Fig.~\ref{fig:integration_curve}. GGC allows for a large reduction of the systematic uncertainties due to the initial kinematics uncertainties, such that $\delta \eta = 10^{-15}$ may be readily achieved within hours of measurement.  This overcomes one of the major challenges for missions like STE-QUEST \cite{Aguilera14} by relaxing the requirements on the source preparation by three orders of magnitude ($\mu$m displacement uncertainty instead of nm, similar for velocity). Even more, the systematics are integrated below $10^{-17}$ within a week and  even reach $10^{-18}$ in a few months. Ultimately, in order to reach these inaccuracies in the \Eotvos\ parameter, the combination of GGC and signal demodulation is indispensable. 

\subsection{Co-location feasibility}
Atom interferometry for metrological applications has enjoyed a surge of interest in the last years \cite{Geiger20,Bongs19}.
In particular, parabolic flights \cite{Barrett16,Geiger11}, drop towers \cite{Vogt20,Condon19,Muentinga13}, sounding rockets \cite{Becker18} and the international space station \cite{Frye19,Elliott18} enable research on atom optics in microgravity including the demonstration of atom interferometry, BEC production, and BEC interferometry in this environment.
In the following, we provide an assessment of the aspects related to the dual-source preparation, in particular the co-location in position and velocity, and evaluate the feasibility of the GGC scheme. 
A complete error model and other aspects of a full space mission are beyond the scope of this paper and are discussed elsewhere \cite{Aguilera14,Williams16,Berge19}.

\paragraph{BEC source} The production of BECs with $10^6$ rubidium atoms in a few seconds are within the capabilities of current devices \cite{Becker18,Hardman16,Rudolph15,Dickerson13}.
Magnetic and optical collimation of the matter waves to 100\,pK and below was demonstrated \cite{Abend16,Kovachy15,Muentinga13}, supporting high beam splitting efficiency \cite{Abend16,Szigeti12} and extended free evolution times \cite{Dickerson13,Muentinga13}.
Mixtures of condensed rubidium and potassium were generated by exploiting Feshbach resonances \cite{Thalhammer08,Ferrari02}, but reaching sufficient numbers of atoms and collimation of both overlapped ensembles requires additional research efforts \cite{phdCorgier}.

\paragraph{Atom interferometry} Beam splitters based on double diffraction providing the required momentum transfer were implemented in interferometric measurements \cite{Gebbe19,Ahlers16,Leveque09} and extended free fall times on the order of seconds were utilised to boost the sensitivity \cite{Dickerson13}. 
Furthermore, the experimental implementation of the GGC scheme has been shown via adjusting the effective wave vector of the central beam splitting pulse and rotation of the mirror \cite{Overstreet18,Damico17,Asenbaum20arxiv}.
In trapped ensembles, Bloch oscillations and the signal of an atom interferometer were observed for total evolution times of up to 20\,s, but with a significantly reduced contrast~\cite{Xu19,Poli11}.
Although interleaved operation has previously been demonstrated in a rotation sensor using a single species \cite{Savoie18}, the dual-species, microgravity operation will require adaptations for the transfer to the interferometer zone \cite{Trimeche_2019} and for assuring the initial overlap.

\paragraph{Requirements on beam splitting light fields imposed by GGC} Current tip-tilt mirror technology to adjust the beam pointing appears to fulfill the requirement stated in Tab.~\ref{tab:params} (see Appendix~\ref{app:realization} for details) since it was utilized to compensate for Earth's rotation with a performance of $1\,\mathrm{nrad/}\sqrt{\mathrm{Hz}}$ \cite{Dickerson13} and repeatable to $\lessapprox1\,\mu\mathrm{rad}$ \cite{Hauth13,Lan12}.
The implementation of the GGC scheme will likely require two lasers per species, where each laser provides two frequencies.
Here, one laser drives the initial and final beam splitter, the second laser provides the central beam splitting pulse with a different and variable wave vector.
In order to ensure the necessary phase stability of the lasers with respect to each other, a reference provided by a frequency comb or a high-finesse transfer cavity is mandatory \cite{phdResch}.
A setup based on only a single laser per species might be possible using fiber lasers offering sufficiently large tuning range.
As a fallback option, the requirement on the tuning range may be relaxed by trading off free fall time against higher beam splitting order \cite{Roura17}.

\paragraph{Satellite platform} Due to its similarity in scope and technological requirements on the satellite platform, the heritage of MICROSCOPE \cite{Touboul17} is essential for the discussion of potential UFF test scenarios. The orbit assumed in this paper is motivated by MICROSCOPE's highly circular orbit at \SI{700}{km} resulting from a trade-off to maximize the local value of $\vec g$ and to minimize atmospheric drag. In particular, the mission has demonstrated excellent attitude and satellite position control \cite{Touboul19} far beyond the parameter assumptions made here in Tab.~\ref{tab:params}. Even better control has been demonstrated in the context of space-borne gravitational wave detection \cite{Armano19}, which is, however, not required for the scenario under consideration.

\section{Conclusion}\label{sec:conclusion}
In this paper, we have illustrated the co-location problem of the two test species in a space-borne quantum test of the UFF way beyond the state-of-the-art. We particularly presented a dual strategy based on variable wave vector shifts and demodulation to mitigate systematic contributions linked to errors in the source preparation. 
Whilst an exhaustive discussion of all sources of noise and systematic effects is beyond the scope of this paper, we have demonstrated that those related to initial co-location uncertainties can be reduced to below $\delta \eta = 10^{-17}$ for realistic experimental scenarios and reasonable mission durations.
At the same time, the requirements on the initial overlap in position and velocity of the two employed species are reduced by five orders of magnitude.
The described methods allow to significantly decrease the required mission duration in proposals like \cite{Aguilera14} and pave the way for missions with unprecedented accuracy beyond state-of-the-art \cite{Berge19}. 

%==============================================================
\begin{acknowledgements}We acknowledge discussions with Holger Ahlers, Robin Corgier, Pac\^{o}me Delva, Florian Fitzek, Christine Guerlin, Thomas Hensel, H\'el\`ene Pihan Le-Bars, Albert Roura, Etienne Savalle, Jan-Niclas Siem\ss, Christian Ufrecht and \'Etienne Wodey. Moreover, we acknowledge financial support from DFG through CRC 1227 (DQ-mat), projects B07 and A05. The presented work is also supported by CRC 1128 geo-Q and the German Space Agency (DLR) with funds provided by the Federal Ministry of Economic Affairs and Energy (BMWi) due to an enactment of the German Bundestag under Grant No. 50WM1641 and 50WM2060. Furthermore, we acknowledge financial support from ''Nieders\"achsisches Vorab'' through the ''Quantum- and Nano- Metrology (QUANOMET)'' initiative within the project QT3 and through ''F\"orderung von Wissenschaft und Technik in Forschung und Lehre'' for the initial funding of research in the new DLR-SI Institute. Moreover, this work was funded by the Deutsche Forschungsgemeinschaft (DFG, German Research Foundation) under Germany’s Excellence Strategy – EXC-2123 QuantumFrontiers – 390837967. S.L. wishes to acknowledge IP@Leibniz, a program of Leibniz Universit\"at Hannover promoted by the German Academic Exchange Service and funded by the Federal Ministry of Education and Research.
D.S. gratefully acknowledges funding by the Federal Ministry of Education and Research (BMBF) through the funding program Photonics Research Germany under contract number 13N14875.

\end{acknowledgements}

%=============================================================

%\bibliography{bib/bibliography}
%

%==============================================================
\appendix
\section{Gravity model}\label{app:model}
For Lagrangians up to quadratic order in $\vec r$ and $\dot{\vec r}$, the interferometer phases may be inferred by a semi-classical model, in which the classical trajectories of the atoms are computed and inserted into the phase expression \eqref{eq:phi}. Throughout this manuscript, we suppose
\begin{equation}\label{eq:L_app}
L = \frac 1 2 m\left( \dot{\vec{r}} + \vec\Omega_s\times\vec r \right)^2 + \frac 1 2 m \vec r \Gamma(t) \vec r,
\end{equation}
to describe the free motion of the atoms in the satellite reference frame.
The local gravity gradient tensor $\Gamma(t)$ depends on the the satellite position and attitude and is therefore a function of time. $\vec \Omega_s$ incorporates rotations of the satellite, i.e. spinning around its own axis.
To this end, we expand the gravitational potential of the Earth for coordinates $\vec r$ much smaller than the satellite position $\vec{R}$, 
\begin{equation}\label{eq:gravpot}
    \begin{aligned}
    \phi(\mathbf{R+r}) &\approx \phi(\mathbf{R})+\partial_{R_i}\phi(\mathbf{R})r_i+\frac{1}{2!}\partial_{R_j}\partial_{R_j}\phi(\mathbf{R})r_ir_j + ... \\ &= \phi(\mathbf{R}) + \mathbf{g}\cdot\mathbf{r} + \frac{1}{2!}\mathbf{r} \Gamma \mathbf{r} +... \end{aligned}
\end{equation}
with
\begin{equation}\label{eq:gamma}
\Gamma_\chi = \begin{pmatrix}
T_{xx} 	& 0 	& T_{xz}\\
0		& T_{yy}& 0		\\ 
T_{xz}	& 0		& T_{zz}   
\end{pmatrix},
\end{equation}
supposing the orbital motion to be restricted to the $x$-$z$-plane. This approximates the potential for the order-of-magnitude assessment performed in this work. A concise mission analysis would involve a realistic gravitational model such as \cite{Pavlis12}. For the Newtonian gravitational potential, the gradient components are given by
\begin{equation}\label{eq:Gamma_ij}
    T_{ij} = \frac{3Gm_E}{|\mathbf{R}|^5}R_iR_j-\frac{Gm_E}{|\mathbf{R}|^3}\delta_{i,j}.
\end{equation}
The parameter $0\leq \chi < 2\pi$, which parametrizes the satellite position on the orbit, is chosen such that the initial position
\begin{equation}
\vec{R}(\chi=0)=(0,0,R_0)
\end{equation}
is aligned with the $z$-axis and the initial gradient tensor reads
\begin{equation}
\Gamma_0 = \begin{pmatrix}
-\gamma/2 	& 0 			& 0				\\
0			& -\gamma/2		& 0				\\ 
0			& 0				& \gamma  
\end{pmatrix},
\end{equation}
with $\gamma = 2GM/R_0^3$.  The way how the time dependent tensor components $T_{ij}$ relate to those of the initial gradient tensor $\Gamma_0$ depends on the shape of the orbit. For a circular orbit (index c), where the satellite position is given by
\begin{equation}
\vec{R}^\text{e}(\chi) = \begin{pmatrix}
R_0 \sin\chi \\ 0 \\ R_0 \cos\chi
\end{pmatrix},
\end{equation}
the explicit relations are
\begin{equation}\label{eq:GG_circ}
\begin{aligned}
T^\text{c}_{xx} &= \frac{1}{4}\gamma\left(1-3\cos(2\chi)\right) \\
T^\text{c}_{zz} &= \frac{1}{4}\gamma\left(1+3\cos(2\chi)\right) \\
T^\text{c}_{xz} &= \frac{3}{4}\gamma\sin(2\chi)
\end{aligned}.
\end{equation}
It is important to note that the modulation of the components is at twice the orbital frequency. In the case of a circular orbit, we can describe the time evolution of the gradient tensor during the interferometer sequence by another rotation, such that it is given by 
\begin{equation}
\Gamma(t) = D(\Omega_m t) \Gamma_\chi D^T(\Omega_m t)
\end{equation}
at time $t$ after the measurement has been started at orbital position $\chi$. $D(\theta)$ is the 3D-rotation matrix by an angle $\theta$ around the $y$-axis.
For an orbit featuring an eccentricity $e$ and semi-major axis $a$, the satellite position is given by \cite{Montenbruck12}
\begin{equation}
\vec{R}^\text{e}(\chi) = \begin{pmatrix}
  R_0 \sqrt{1-e^2}\sin\chi\\ 0 \\R_0 (\cos\chi-e).
\end{pmatrix}
\end{equation}
Note that the initial position coincides with perigee, $R_0 = a(1-e)$ in the coordinate system fixed to center of the Earth. With the help of \eqref{eq:Gamma_ij}, we readily obtain
\begin{equation}
\begin{aligned}
T^\text{e}_{xx} &=  -\frac{1-4e^2+4e\cos(\chi)+(2e^2-3)\cos(2\chi)}{4|1-e\cos\chi|^5/|1-e|^3}\ \gamma\\
T^\text{e}_{zz} &=  -\frac{1-3e^2+4e\cos(\chi)+(e^2-3)\cos^2\chi}{2|1-e\cos\chi|^5/|1-e|^3}\ \gamma\\
T^\text{e}_{xz} &= -\frac{3\sqrt{1-e^2}(\cos \chi-e)\sin\chi}{2|1-e\cos\chi|^5/|1-e|^3}\ \gamma
\end{aligned}
\end{equation}
A series expansion to first order in the ellipticity yields
\begin{equation}\label{eq:GG_ell}
\begin{aligned}
	T^\text{e}_{xx} &= T^\text{c}_{xx}+\frac{3}{8}\gamma \left[-2+\cos(\chi)+6\cos(2\chi)-5\cos(3\chi)\right]e \\
	T^\text{e}_{zz} &= T^\text{c}_{zz}+\frac{3}{8}\gamma \left[-2+3\cos(\chi)-6\cos(2\chi)+5\cos(3\chi)\right]e \\
	T^\text{e}_{xz} &=	T^\text{c}_{xz}+\frac{3}{8}\gamma \left[\sin(\chi)-6\sin(2\chi)+5\cos(3\chi)\right]e 
\end{aligned}
\end{equation}
which shows that the ellipticity introduces an additional modulation of the gradient components at different frequencies compared to the circular orbit \eqref{eq:GG_circ}. In particular, it features a component at the orbital frequency, which leads to a systematic contribution even after demodulation  ($\Delta a_\text{sys}^1$ in Eq.\,\eqref{eq:avg}), since is modulated at the same frequency as a possible UFF violation signal. However, it is suppressed by $e$ and can be accounted for with the compensation technique of section \ref{sec:ggc}. Similarly, higher orders in the expansion of the gravitational potential \eqref{eq:gravpot} feature frequency components at the orbital frequency. However, the associated acceleration uncertainties are suppressed with respect to the gravity gradient terms discussed in this paper by a factor of $\Delta r/|\vec R|\sim 10^{-12}$.

\section{Implementation and feasibility of the compensation method}\label{app:realization}
\subsection{Experimental method}
The shifts $\Delta_{x,j}$, $\Delta_{z,j}$ in the wave vector of the $j$-th pulse, required to compensate the gravity gradient induced acceleration uncertainties as outlined in section \ref{sec:ggc} in the main text, are realized by tilting the laser by an angle $\theta_j$ and shifting it in frequency by $\Delta f_j$. We find the relations
\begin{equation}\label{eq:tiltshift}
    \begin{aligned}
        \theta_j      &= \arctan\left(\frac{\Delta_{x,j}}{1+\Delta_{z,j}}\right)\\
        \Delta f_j    &= \frac{c\keff}{8\pi}\times \left[-1+\sqrt{(\Delta_{x,j})^2+(1+\Delta_{z,j})^2}~\right],
    \end{aligned}
\end{equation}
where the factor 8 accounts for two-photon transitions employed to realize the beam splitters and the second order diffraction process. $c$ is the speed of light and $\keff = |\vec\keff| = 4k_L$ the effective total momentum transferred by the pulses, with $k_L$ being the wave number of the light.

\subsection{Treatment of residual rotations}
The employed Lagrangian \eqref{eq:L} with $\vec\Omega_s=\vec 0$ describes the perfectly inertial case, i.e. the constant rotation of the gravity gradient tensor by an angle $\Oo t$. Residual rotations of the satellite, however, modify the Lagrangian in two ways. In the following we assume the same rotation uncertainty $\delta \Omega$ for all three directions which gives rise to 
\begin{equation}\label{eq:L_delOm}
L_{\delta \Omega} = \frac 1 2 m\left( \dot{\vec{r}} + (\delta\Omega,\delta\Omega,\delta\Omega)\times\vec r \right)^2 + \frac 1 2 m \vec r D_\dO^T \Gamma_\chi D_\dO \vec r,
\end{equation}
where $D_\dO$ is the matrix that rotates the gradient tensor under consideration of the orbital motion and residual rotations. 
Under the assumption that $\dO t << 1$ (the case here), for an arbitrary permutation $D_p = [D_{\delta \Omega,x}D_{\delta \Omega,y}D_{\delta \Omega,z}D_{\Oo,y}] $  of these four rotations (three small residual rotations, one comparably large orbital rotation), the relative deviation of the term $D_p^T \Gamma_\chi D_p$ from $D_{\Oo}^T \Gamma_\chi D_{\Oo}$ is smaller than $10^{-6}$ for the parameters of interest. Therefore, neglecting the residual rotations results in an error of less than $10^{-11}$\,s$^{-2}$ in the value of the gradient tensor components, which is well in line with the assumptions. As a consequence, the impact of residual rotations on the modulation of the gradient tensor can be safely neglected, and the Lagrangian
\begin{equation}
L_{\delta \Omega} = \frac 1 2 m\left( \dot{\vec{r}} + \vec {\delta \Omega}\times\vec r \right)^2 + \frac 1 2 m \vec r D_{\Oo}^T \Gamma_\chi D_{\Oo} \vec r
\end{equation}
with $\vec{\delta \Omega}=(\delta\Omega,\delta\Omega,\delta\Omega)$ captures the relevant influence of residual rotations on the system.

\subsection{Uncertainty assessment}\label{app:unc}
Due to the linear dependence on the initial kinematics, the differential acceleration between two species $A$ and $B$,
\begin{equation}\label{eq:a_diff}
\Delta a = a_A-a_B = \Delta a_\text{indep}+\sum_{i=1}^3 \left(\alpha'_i r_{0,i}  + \beta'_i v_{0,i}\right),
\end{equation}
can be directly obtained from the phase expression \eqref{eq:phi_dep} after division by the respective scale factor $k_{\text{eff},\alpha} T_\alpha^2$ of species $\alpha$, i.e. $\alpha'_i = \alpha_{i,A}/k_{\text{eff},A}T_A^2-\alpha_{i,B}/k_{\text{eff},B}T_B^2$ (ana-logous for $\beta'_i$). The total differential acceleration uncertainty is given by the absolute sum 
\begin{equation}\label{eq:del_a}
\begin{aligned}
 \delta \Delta a =&  \left|\delta \Delta a_\text{indep}\right|\\
 &+\sum_{i=1}^3 \Big(|\delta \alpha'_i|\Delta r_{0,i}  + (|\alpha'_i|+|\delta\alpha'_i|) \delta \Delta  r_{0,i}\vphantom{\sum}\\
 &\phantom{+\sum_{i=1}^3 \Big(} |\delta \beta'_i| \Delta  v_{0,i} +(|\beta'_i|+|\delta\beta'_i|) \delta \Delta v_{0,i}\Big),
\end{aligned}
\end{equation}
where $\delta$ indicates the uncertainty of a term. For example, $\delta \Delta r_{0,i}$ is the uncertainty in the initial displacement $\Delta r_{0,i}$. The error in the coefficients $\alpha'_i$ is computed via
\begin{equation}
\delta\alpha'_i = \sum^N_{j=1}\left|\frac{\partial \alpha'_i}{\partial Q_j}\delta Q_j\right|,
\end{equation}
with $Q_j \in \{\gamma, T, \vec\Omega_s, \Delta_{2,x},\Delta_{2,z},\Delta_{3,x},\Delta_{3,z}\}$ and $\delta Q_j$ being the corresponding uncertainty (analogous for $\beta'_i$). Note that this is a conservative treatment, since most of these contributions are uncorrelated such that the favourable quadratic sum would be sufficient. To first order, $\Delta_{x,j} \sim \theta_j$ and $\Delta_{z,j} \sim \Delta f_j/f$ such that their uncertainties are given by $\delta \Delta_{x,j} \sim \delta\theta_j$ and $\delta \Delta_{z,j} \sim \delta \Delta f_j/f$, respectively. 

The differential acceleration uncertainty of a measurement without the compensation technique is obtained from \eqref{eq:del_a} with $\Delta_{i,j} = \delta \Delta_{i,j} = 0$. By construction, application of the compensation shifts $\Delta_{i,j}$ derived in  Sec.\,\ref{sec:ggc} leads to $\alpha'_i = \beta'_i = 0$ in Eq.\,\eqref{eq:del_a}, i.e.
\begin{equation}\label{eq:del_a_GGC}
\begin{aligned}
\delta \Delta a_\text{GGC}(t) =& |\delta a_\text{indep}|\\
& + \sum_{i=1}^3\Big( |\delta \alpha'_i| \left(\Delta r_{0,i} +\delta \Delta r_{0,i} \right) \\
& \phantom{+ \sum_{i=1}^3\Big(}+ |\delta \beta'_i| \left(\Delta v_{0,i} + \delta \Delta v_{0,i}\right)\Big).
\end{aligned}
\end{equation}

As discussed in Sec.\,\ref{sec:demodulation}, the integrated uncertainty is then given by
\begin{equation}\label{eq:intunc}
    \delta a(\tau) = \frac 2 \tau \int^\tau_0 \delta \Delta a_\text{GGC}(t) \cos(\Oo t) \dt.
\end{equation}

\subsection{Demodulation of a discrete sample}
In order to account for the finite sampling of the data due to the experimental cycle time $T_c$, the averaging expression \eqref{eq:intunc} needs to be discretized,
\begin{equation}
    \delta a(n) = \frac{2}{n} \sum_{m=1}^n \delta \Delta a_\text{GGC}(m T_c) \cos(\Oo m T_c) .
\end{equation}
The total integration time corresponding to $n$ measurements is consequently given by $\tau=nT_c$.

\subsection{Noise}\label{app:noise}
The integration of noise is modified in the presence of a modulated signal. For an order-of-magnitude assessment, we refer to a simplified model in which the time signal of a differential measurement is given by
\begin{equation}
    \Delta a = \eta g_0 \cos(\Oo t) + \sigma_a,
\end{equation}
with $\sigma_a$ being the atomic shot noise. The covariance of $\eta$ after $n$ measurements is quantified by
\begin{equation}
    \text{cov}(\eta) = \sigma^2 \left(\vec X^T \vec X \right)^{-1},
\end{equation}
where $X_m = g_0\cos(\Oo m T_c)$ captures the modulated local value of the gravitational acceleration in the $m^\text{th}$ measurement, $T_c$ is the cycle time. Consequently, the statistical uncertainty due to shot noise in the determination of $\eta$ is given by
\begin{equation}
\begin{aligned}
\sigma_\eta(n) &= \sqrt{\text{cov}(\eta)} \\
&= \frac {\sigma}{g_0} \left( \sum_m^n \cos^2(\Oo m T_c)\right)^{-1} \\
&\rightarrow \frac{\sigma \sqrt 2}{g_0 \sqrt n}
\end{aligned}
\end{equation}
in the limit of many measurements, $\tau = n T_c\gg T_c$.

\end{document}